# GAMORRA: An API-Level Workload Model for Rasterization-based Graphics Pipeline Architecture

Iman Soltani Mohammadi[1], Mohammad Ghanbari[2], Life Fellow, IEEE and Mahmoud Reza Hashemi[3]

## Abstract

The performance of applications that require frame rendering time estimation or dynamic frequency scaling, rely on the accuracy of the workload model that is utilized within these applications. Existing models lack sufficient accuracy in their core model. Hence, they require changes to the target application or the hardware to produce accurate results. This paper introduces a mathematical workload model for a rasterization-based graphics Application Programming Interface (API) pipeline, named GAMORRA, which works based on the load and complexity of each stage of the pipeline. Firstly, GAMORRA models each stage of the pipeline based on their operation complexity and the input data size. Then, the calculated workloads of the stages are fed to a Multiple Linear Regression (MLR) model as explanatory variables. A hybrid offline/online training scheme is proposed as well to train the model. A suite of benchmarks is also designed to tune the model parameters based on the performance of the target system. The experiments were performed on Direct3D 11 and on two different rendering platforms comparing GAMORRA to an AutoRegressive (AR) model, a Frame Complexity Model (FCM) and a frequency-based (FRQ) model. The experiments show an average of 1.27 ms frame rendering time estimation error (9.45%) compared to an average of 1.87 ms error (13.23%) for FCM which is the best method among the three chosen methods. However, this comes at the cost of 0.54 ms (4.58%) increase in time complexity compared to FCM. Furthermore, GAMMORA improves frametime underestimations by 1.1% compared to FCM.

## 1. INTRODUCTION

In traditional Cloud Gaming (CG) [1], rendering is performed at the server side while in graphics streaming-based CG [2, 3] all the frames are rendered at client side. As a compromise between these two methods, hybrid graphics/video streaming CG [4, 5] has been proposed in which there is a potential for some frames to be rendered at client side. The latter two CG methods require to estimate the rendering time of each frame to make sure that the client device can handle the workload of the game for every frame that needs to be rendered at client side. In addition, real time collaborative rendering platforms like Kahawai [6] and other similar platforms [7], require a reliable workload model to set the graphical level of details of every frame based on the computational power of the thin client before rendering is carried out. Also, Dynamic Voltage and Frequency Scaling (DVFS)-based power management systems for mobile games, which reduce the power consumption of a processor by dynamically adjusting its voltage and frequency, need to take into account the amount of workload of each frame [8-11]. These studies usually rely on simple workload models based on the number of triangles [12] or mostly focus on predicting the upcoming frames' workload by using a simple linear model [13]. These models try to compensate for their lack of sufficient detail in their core model by operating at hardware level.

Due to the high variety of GPU architectures among vendors, a hardware level model effectively limits the applicability of such a model to a specific hardware device. Although each graphics driver covers a certain range of hardware models, they also differ significantly from each other due to their hardware dependent nature and the frequent updates they receive. But graphics APIs usually follow a certain model for their rendering pipeline with minor differences between different APIs' pipeline architectures. Hence, designing a mathematical model to estimate frame rendering times (frametime) at the graphics API level covers a much wider range of applications compared to a hardware architecture or driver level model. Additionally, other proposed methods [14, 15] usually require changes to the hardware or software to provide an accurate estimation. Utilizing an API-level model avoids the need for game engine modifications or hardware-level changes that are not possible in case of commercial off-the-shelf products. Therefore, a Graphics API-level Model of Rendering workload for Rasterization-based graphics pipeline Architecture, GAMORRA, is proposed in this paper. However, to compensate for this high-level approach which would inevitably lead to loss of accuracy, three components are devised for GAMORRA to ensure an accurate prediction: (i) a detailed regression core model, (ii) a customized training scheme to train the core model weights, and (iii) a suite of benchmarks to tune the model parameters based on the performance of the target hardware.

Modern API pipelines consist of fixed-function and programmable stages as opposed to the fully fixed-function pipelines of old rendering systems. The early experiments to determine the proper core model, as well as the benchmark results in Section 4.1., indicate a linear relation between the performance of each stage of the pipeline and the overall rendering time of each frame. The results of these experiments show that each stage of the pipeline contributes differently to the final rendering time. This difference which stems from the difference in the computational complexity of the stages, despite the same underlying hardware, indicates their independence as explanatory (independent) variables. Also, the linear relation between the performance of each stage and the overall processing time of the pipeline which acts as the response (dependent) variable, suggests that a Multiple Linear Regression (MLR) [16] technique has the potential to serve as a core for the proposed model. Hence, as opposed to previous studies such as [9] that consider only one explanatory variable, in GAMORRA which takes advantage of MLR at its core, an explanatory variable is dedicated to each stage of the graphics API pipeline. This approach provides more flexibility and a better imitation of the rendering process of a rasterization-based application which leads to more accurate frametime estimations.

[1] School of Electrical and Computer Engineering, University of Tehran, Tehran, Iran (e-mail: soltani.m@ut.ac.ir)
[2] School of Electrical and Computer Engineering, University of Tehran, Tehran, Iran (e-mail: ghan@ut.ac.ir) as well as an Emeritus Professor at the School of Computer Science and Electronic Engineering, University of Essex, Colchester, UK (e-mail: ghan@essex.ac.uk)
[3] School of Electrical and Computer Engineering, University of Tehran, Tehran, Iran (e-mail: rhashemi@ut.ac.ir)

Using an offline training scheme to train model weights helps with avoiding the need to train the weights from scratch during runtime which can compromise the real-time functionality of the model especially at the start of a rendering session. However, only a limited number of samples can be used to train the model weights prior to a rendering session, which would not be representative of the whole gameplay. Online training [17] can be used to adapt the weights to the changes during runtime, on the fly. Hence, a new hybrid offline/online training method is proposed to train the model prior to a rendering session (offline) to obtain an acceptable set of initial weight values while updating the weights during runtime (online). This approach also helps with avoiding overfitting which might be experienced in case of an overly complicated model and an extensive offline-only training scheme.

Additionally, to accurately tune the model parameters based on the performance of the target rendering system, a suite of benchmarks is designed to assess the performance of each stage of the pipeline according to GAMORRA's workload model.

In summary, the primary contributions of this study are as follows:

- **A reliable and practical model for frametime estimation of rasterization-based commercially off-the-shelf software and hardware**
- **A hybrid offline/online training scheme to train the proposed model**
- **A benchmark suite to evaluate the performance of the target rendering system and tune the model parameters accordingly**

The rest of the paper is organized as follows. The next section discusses notable works in this field and how GAMORRA differs from them. Section 3 explains the core design and functionality of GAMORRA in detail. Section 4 focuses on the implementation details of the proposed method and the experimental results. And finally, the paper is concluded in Section 5.

## 2. RELATED WORK

A limited number of studies have focused on estimating frametimes by targeting different applications for their proposed methods. Some of the studies in this field require hardware and software changes to perform properly which is not desired in the case of closed source software and already available hardware.

Wimmer et al. [14] proposed to use the number of transformed vertices and the number of projected pixels for each object to estimate frametime. They propose a hardware extension to further improve the estimation accuracy.

Mochocki et al. [18] proposed a signature-based model for workload estimation in which each signature is calculated based on the number of triangles and the transformations that are performed on the vertex data. Focusing on the number of triangles and geometrical transformations is not sufficient for a programmable pipeline in a realistic scenario and results in reduced precision.

Gu et al. [19] proposed a hybrid workload prediction method that switches between a Proportional-Integral-Derivative (PID) [20] controller-based and a frame structure-based prediction scheme. This method is mainly used to predict the workload of an upcoming frame to be used in a Dynamic Voltage Scaling (DVS) power management system. In this work, rasterization workload is considered as the most significant contributing source of processing time in rendering. Similarly, Zhang-Jian et al. [8] proposed to use the number of triangles as a measure of workload complexity and a PID controller to predict the workload of each frame. Dietrich et al. [13] further expanded PID-based methods and proposed to use the Least Mean Squares (LMS) method for controller parameter identification to avoid the need to hand-tune the parameters.

Dietrich et al [21] also proposed to predict each frame's workload using an autoregressive model by considering the previous frames' number of cycles as the explanatory variable. They further expanded their work by proposing a self-tuning LMS linear predictor to estimate the parameters of an AutoRegressive (AR) moving average model for workload (number of cycles) prediction [9]. Solely relying on the previous frames' workload without taking the characteristics of the frame into account is severely misleading due to the heavily variant workload of graphical scenes even in consecutive frames. These methods usually fail to react in time to the workload variations of frames.

Cheng et al. [22] proposed a behavior-aware power management system for mobile games which estimates each frame's workload based on the number of game application's API calls and texture processing load.

Song et al. [23] proposed a fine-grained GPU power management called Frame Complexity Model (FCM) for closed source mobile games which works based on the number of vertices, the number of API commands and the size of textures. This approach does not take the impact of other contributing factors such as the complexity of shader programs or the structure and performance of the graphics API that processes all the aforementioned data, into account. This causes such a model to produce the same results for different APIs.

Gupta et al. [10] proposed a light-weight adaptive runtime performance model to estimate the sensitivity of frametimes to the current GPU frequency. This method consists of two steps: first, an offline data collection process is performed where the required data on frametimes and GPU performance counters are collected. Second, the collected data are used to tune a differential frametime model and predict the frametimes. This method considers the overall frametime of the previous frames to predict future frametimes based on the changes that are made to the GPU frequency and GPU counters. Also, an online learning scheme is employed to update the parameters at runtime. In this work, the graphical structure of a frame is ignored. Also, GPU counter values are unknown when a batch [24] is not yet processed which can be problematic for practical use.

Cheng et al. [15] proposed to use the changes in frequency and the number of active GPU computational slices for power management in mobile games. This method requires to be implemented in GPU firmware to achieve sufficient accuracy.

Choi et al. [25] proposed a predictive method for frametime estimation based on previous frametimes and the frequency at which they were rendered. This study, similar to other more recent studies [26], focuses on big.LITTLE architecture in mobile devices.

To summarize, some of the proposed methods on workload modeling require modifications in the application or the target hardware [14, 15, 27]. GAMORRA attempts to avoid such requirements by operating at an API-level. In order to compensate for its high-level approach, GAMORRA considers the workload of all the stages of the pipeline and the pipeline architecture, unlike numerous studies [8, 10, 11, 19, 28] that focus on a limited number of contributing factors. Additionally, as opposed to multiple studies that are designed for a specific hardware [25, 26], GAMORRA is independent of the underlying architecture.

## 3. PROPOSED MODEL

GAMORRA acts as a middleware that resides between the application and the graphics API software, capturing the output API commands produced by the application's rendering engine. Figure 1 shows the placement of GAMORRA in a computer system. GAMORRA analyzes the graphics data stream to obtain the value of the contributing factors to the workload so that they are fed to the MLR model as the explanatory variables. The overall workload of the model and the workload of each stage is discussed in subsection 3.1. Then, the training process is explained in subsection 3.2 followed by some notes on the benchmark suite in subsection 3.3.

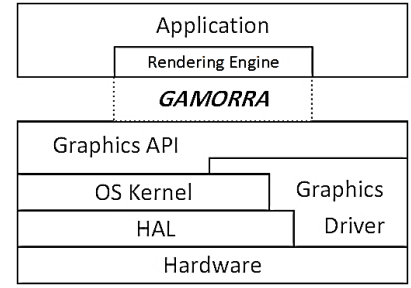

Figure 1 GAMORRA's placement in a rendering system

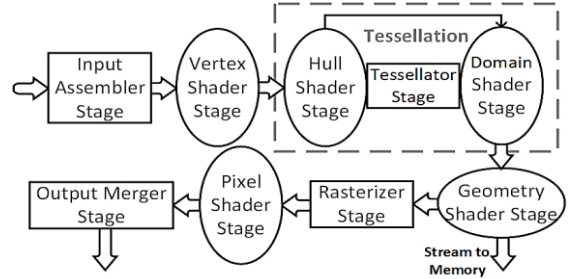

Figure 2 The overall architecture of the Direct3d 11 pipeline

### 3.1. Workload model

The overall architecture of a modern graphics API pipeline (Direct3D 11 in this case) is shown in Figure 2. Direct3D 11's graphics pipeline consists of 4 fixed-function stages (marked by rectangle containers) and 5 shader stages (marked by oval-shaped containers), a total of 9 stages: Input Assembler (IA), Vertex Shader (VS), Hull Shader (HS), Tessellator Stage (TS), Domain Shader (DS), Geometry Shader (GS), Rasterizer (Ras), Pixel Shader (PS), and Output Merger (OM).

A frame is broken down into multiple rendering batches. Each batch has a different pipeline state that result in one or more *drawcall*s which draw the pixels prepared by the current batch [24]. Batches are rendered sequentially. Each batch can contain multiple drawcalls as long as these drawcalls do not cause any state changes to the pipeline. Since there are no state changes in this case, GAMORRA considers the drawcalls in a batch as a single drawcall. Hence, the proposed frametime model formulates the overall estimated processing time of the $i^{th}$ frame, $T_i$, as in (1) where $B_i$ is the number of batches for the $i^{th}$ frame and $Batch_b$ shows the estimated processing time of the $b^{th}$ batch:

$$T_i = \sum_{b=0}^{B_i-1} Batch_b \qquad (1)$$

In the case of Direct3D 11, there are 9 explanatory variables required which are represented by $w_n^b$ for the $b^{th}$ batch with $n = 1, \ldots, 9$. Let $\beta_n$ represent the model parameter of each stage with $n = 1, \ldots, 9$, $\beta_0$ represent the minimum processing time of the pipeline and $\varepsilon$ be the overall model error, then the MLR model of the rendering pipeline is defined as:

$$Batch_b = \beta_0 + \sum_{n=1}^{9} w_n^b \beta_n + \varepsilon \tag{2}$$

To simplify (2), $w_0^b$ is defined and set to 1 for all the batches. Then (2) can be written as:

$$Batch_b = \sum_{n=0}^{9} w_n^b \beta_n + \varepsilon = \beta w + \varepsilon \tag{3}$$

where $W$ and $\beta$ are vectors containing explanatory variables of the model and model parameters respectively which are defined as:

$$w = [1, w_1^b, w_2^b, w_3^b, w_4^b, w_5^b, w_6^b, w_7^b, w_8^b, w_9^b]^T$$
$$\beta = [\beta_0, \beta_1, \beta_2, \beta_3, \beta_4, \beta_5, \beta_6, \beta_7, \beta_8, \beta_9] \tag{4}$$

Since $w$ is actually a $1 \times 10$ matrix, the $T$ superscript in the above equation represents the transpose operation. The values in $\beta$ are obtained by using the Singular Value Decomposition (SVD) [29, 30]. But before obtaining the model parameters, the explanatory variables $W$ should be known. These variables represent the load of each stage in time unit and they are calculated as discussed in the next sub-section.

### 3.1.1. Stage workload

To model the workload of each stage, a generic formula is proposed. The formula is able to incorporate the details that may differ in various stages of the pipeline due to their unique characteristics. Let us assume that $Perf_n$ represents the performance function of the $n^{th}$ stage in time unit, $L_n^b$ stands for the load of the $n^{th}$ stage and $\eta$ determines the number of cores of the GPU, then the workload of each stage is modeled as:

$$w_n^b = Perf_n(L_n^b)/\eta \tag{5}$$

$Perf_n$ maps the amount of load of the $n^{th}$ stage which is the number of input elements (number of vertices or pixels) to the performance weight of the stage for that specific amount of load. The value of this performance weight roughly estimates the overall processing load of each stage compared to the other stages of the pipeline. $Perf_n$ is determined by the custom graphics benchmarks that are designed specifically for GAMORRA and are discussed in Section 3.2.

For a fixed function stage, $L_n$ mainly depends on the number of inputs or outputs of that stage. But, the overall processing load of shaders is strongly affected by their shader program as well. The shader compilation is performed in 2 stages: first a tool compiles the HLSL code into the GPU agnostic Intermediate Language (IL). Then the GPU driver converts the IL into the final shader assembly (ISA) that can be executed on a specific GPU. The complexity of the shader program is obtained through analyzing its Intermediate Language (IL) assembly code which comprises a series of instructions each of which performs a specific operation based on its opcode. To obtain the overall time complexity of a shader, the time complexity of each IL opcode needs to be determined through benchmarking. Let the number of assembly operators be shown by $N_{op}$, the processing time of the $j^{th}$ operator be represented by $op_j$, the number of its occurrences in the current shader be represented by $x_j$ and $N_i$ represent the number of times that the shader is invoked, then the complexity of a programmable stage, $C_i$, and consequently, $L_i$ is calculated as follows:

$$C_n^b = \sum_{j=0}^{N_{OP}-1} op_j^n . x_j^n$$

$$L_n^b = \begin{cases} N_n^b, & \text{if fixed function} \\ C_n^b . N_n^b, & \text{if Programmable} \end{cases} \tag{6}$$

It should be noted that the shader complexity model is not meant to act as an accurate stand-alone model for shader processing time estimation. The goal of this model is to provide the core model of GAMORRA with a rough estimate of the overall complexity of each shader and still perform with enough accuracy while ensuring real-time functionality of the overall model. Since shaders are invoked per input, $N_i$ represents the number of vertices for VS, HS, DS, and GS and it represents the number of pixels for PS. The IA stage reads and prepares the vertex data that are required for the current batch by determining their attributes and topology. For the IA stage, since the data is read from resource buffers, the available memory bandwidth becomes the potential bottleneck. Hence, the load of the IA stage ($L_{IA}$) is chosen to be the size of the input vertex data which might vary based on the number of vertices and their attributes (#Attr).

Since Vertex Shader (VS) is a programmable stage, its performance depends on the complexity of its code and should be reflected in $L_{VS}$. VS program is invoked individually for each vertex, so $L_{VS}$ is also affected by the number of vertices. Let $C_{VS}$ be the assembly code's complexity, then $L_{VS}$ is defined as:

$$L_{VS} = C_{VS} . N_{Vertex} \tag{7}$$

For shaders, the assembly operators are profiled separately and a performance function is derived for each operator. Some operators are used exclusively in a specific stage (e.g., sampling operator for PS).

Tessellation stages consist of three separate stages, HS, the Tessellator and DS that act as a single unit and turning off tessellation disables all the underlying stages. These stages are programmed differently but they all work together and they manipulate the vertices in a *patch*. HS consists of a main shader program and a patch constant function (PCF) that are executed once per *output* control point and once per patch respectively. Hence, the load of this stage can be simply considered as the number of vertices ($N_{Vertex}$) along with the number of patches ($N_{PCF}$) as the input to the PCF which is treated like a complete shader stage. The complexity of HS's main shader and PCF are shown by $C_{HS}$ and $C_{PCF}$ respectively while the load for each one is represented by $L_{HS}$ and $L_{PCF}$. $Perf_{HS}$ and $Perf_{PCF}$ are calculated as:

$$L_{HS} = C_{HS} . N_{Vertex}$$
$$L_{PCF} = C_{PCF} . N_{PCF}$$
$$Perf_{THS} = Perf_{HS}(L_{HS}) + Perf_{PCF}(L_{PCF}) \tag{8}$$

The Tessellator stage is also a fixed function unit and its inputs are the tessellation factors and patch constant data that are produced by PCF. For this stage, $L_{Tess}$ mainly depends on the total number of newly generated points in each patch where the total number of patches is shown by $P$ and the number of tessellations in the $p^{th}$ patch by $N_{Tess}^p$:

$$L_{Tess} = \sum_{p=0}^{P-1} N_{Tess}^p \qquad (9)$$

DS stage is fed with the output of the Tessellator and HS stages, namely UVW coordinates of every point in a patch from the Tessellator along with control points and patch constants from the HS stage. The DS stage produces a single tessellated vertex per input vertex, so $L_{DS}$ depends on the number of vertices as well as the complexity of DS's code, $C_{DS}$, and is calculated as:

$$L_{DS} = C_{DS}.N_{DS} \qquad (10)$$

GS is an optional stage which handles complete primitives instead of a single vertex. In addition to the complexity of GS's code, $C_{GS}$, the load of this stage, $L_{GS}$, is also dependent upon the number of vertices which might be different from the input of VS due to being processed in tessellation stages (if tessellation is on). Hence, $L_{GS}$ is calculated as:

$$L_{GS} = C_{GS}.N_{GS} \qquad (11)$$

Rasterizer generates fragments that might end up on screen as pixels. Hence, $L_{Ras}$ which represents the load of the Rasterization stage is considered to be equal to the number of fragments that are produced in this stage, $N_{Frag}$.

PS or fragment shader is the last programmable stage that manipulates each input fragment's color [31]. As the number of fragments $N_{Fragment}$ that are produced by the Rasterizer increases, the number of times that a PS is invoked increases as well. Also, the PS code complexity $C_{PS}$ should be considered in the model. hence, $L_{PS}$ is calculated as:

$$L_{PS} = C_{PS}.N_{Fragment} \qquad (12)$$

OM is the last stage of the pipeline and the fragments that are processed by the PS are fed to this stage. Reading and writing to the render targets are the main cause of the performance issues related to the OM stage when blending is utilized. Hence, this stage is mostly bandwidth limited and is affected by the number of fragments, $N_{Fragments}$ along with the render target resolution, $N_{Width} \times N_{Height}$. $L_{OM}$ is defined as:

$$L_{OM} = N_{Width}.N_{Height}.N_{Fragments} \qquad (13)$$

Compute Shader (CS) is considered as a tool for General-purpose computing on GPUs (GPGPU). This shader has an independent logical pipeline dedicated to general computations. Although CS is not part of the main pipeline and it is absent in Figure 2, this shader should be considered in the performance model, following the same rule for input size, $N_{Input}$, and code complexity, $C_{CS}$. Hence, the load of CS, $L_{CS}$, is calculated as:

$$L_{CS} = C_{CS}.N_{Input} \qquad (14)$$

### 3.1.1. Model parameters

The parameter estimation method aims to minimize the least squares problem produced by the $M$ observations that are recorded by the benchmark. Considering (2), let $y_m$ represent the actual output of the $m^{th}$ observation in the benchmark and $\varepsilon_m$ represent the estimation error for the $m^{th}$ observation, then the following linear system is obtained:

$$\begin{cases} y_0 &= \beta_0 + \beta_1 w_1^0 + \ldots + \beta_9 w_9^0 + \varepsilon_0 \\ y_1 &= \beta_0 + \beta_1 w_1^1 + \ldots + \beta_9 w_9^1 + \varepsilon_1 \\ &\vdots \\ y_{M-1} &= \beta_0 + \beta_1 w_1^{M-1} + \ldots + \beta_9 w_9^{M-1} + \varepsilon_{M-1} \end{cases} \qquad (15)$$

To obtain the matrix form of the system, similar to (3), let $Y$ be the vector containing $y_m$ values, $W$ be the matrix that holds the explanatory variables values and $E$ be the vector for error values, then the linear system in (17) can be written as:

$$Y = \beta W + E \qquad (16)$$

where

$$Y = [y_0, y_1, \ldots, y_{M-1}]$$

$$W = \begin{bmatrix} w_0^0 & w_0^1 & \cdots & w_0^{M-2} & w_0^{M-1} \\ w_1^0 & w_1^1 & \cdots & w_1^{M-2} & w_1^{M-1} \\ \vdots & \vdots & \ddots & \vdots & \vdots \\ w_8^0 & w_8^1 & \cdots & w_8^{M-2} & w_8^{M-1} \\ w_9^0 & w_9^1 & \cdots & w_9^{M-2} & w_9^{M-1} \end{bmatrix}$$

$$E = [\varepsilon_0, \varepsilon_1, \ldots, \varepsilon_{M-1}] \qquad (17)$$

The solution to this linear system minimizes the L2 norm of $E$ formulated as:

$$\min \|E\|_2^2 = \min \|Y - \beta W\|_2^2 = \min J(\beta_0, \ldots, \beta_9) \qquad (18)$$

where $J(\beta)$ represents the cost function.

SVD uses orthogonal transformations to reduce the problem to a diagonal system. If $U$ is an $M \times M$ orthogonal matrix, $S$ is an $M \times 10$ diagonal matrix and $V$ is a $10 \times 10$ orthogonal matrix, then in this method:

$$W^M = USV^T \qquad (19)$$

Considering the above equation, $\beta$ can be formulated as:

$$\beta = ((USV^T)^T USV^T)^{-1}(USV^T)^T Y = VS^{-1}U^T Y \qquad (20)$$

### 3.2. Training the model

An offline training is performed before each rendering session based on a set of frames that use the graphical data of the current scene that is going to be rendered. An offline-only training scheme can be used for some simpler games that do not have aggressive variations in their workload. However, it is not possible to cover all types of workload variations in offline training for modern games that might have over 100 hours of gameplay and heavily variant and dynamic environments. on the other hand, an online only training scheme can lead to long

**Algorithm 1** Mode decision algorithm

**Input:** Current frametime, predicted frametime using online and offline model weights

1: $n = 0$, $n_v = 0$, $CurrentMode = Offline$
2: Run GAMORRA's benchmark according to
3: **while** $n < N$
4:   **if** $CurrentMode == Online$
5:     Calculate $RMSE_n^{on}$
6:     **if** $RMSE_n^{on} > RMSE_{Th}$
7:       Calculate $RMSE_n^{off}$
8:       **if** $RMSE_n^{on} > RMSE_n^{off}$
9:         $n_v = n_v + 1$
10:         **if** $n_v > patience$
11:           Stop online training
12:           Use offline training weights
13:           $CurrentMode = Offline$
14:           $n_{vi} = 0$
15:         **end if**
16:       **end if**
17:     **end if**
18:   **else if** $CurrentMode == Offline$
19:     Calculate $RMSE_n^{off}$
20:     **if** $RMSE_n^{off} > RMSE_{Th}$
21:       Start online training
22:       $CurrentMode = Online$
23:     **end if**
24:   **end if**
25:   $n = n+1$
26: **end while**

initiation times before a set of acceptable and accurate set of weights is obtained. In such cases, using offline training is beneficial to provide the model with acceptable initial weight values to avoid long initiation times of online training. Additionally, huge variations in consecutive frametimes that are not caused by the workload of the frames can make the estimation error during online training to raise beyond an acceptable value. Resetting the model weights to weights obtained by the offline training in such cases can be beneficial, until the error of the offline mode becomes unacceptable as well. Hence, a hybrid offline/online training technique is used to take advantage of both training methods.

Initially, the model starts in offline mode which uses the weights that were obtained through offline training. If the Root Mean Squared Error (RMSE) value for frame $n$ during the offline mode ($RMSE_n^{off}$) increases beyond a predetermined threshold ($RMSE_{Th}$), the model switches to online mode. The value of $RMSE_{Th}$ determines the amount of acceptable estimation error. The acceptable $RMSE$ value is generally under 0.5 for this metric which is also used as the value for $RMSE_{Th}$ in the experiments as well. However, tuning this parameter per game can have a positive impact on the overall accuracy. Setting the value of this parameter too high reduces the sensitivity of the system to estimation error, while too small values make the system constantly switch between online and offline modes.

Upon switching to the online mode, the online training starts and the model weights, which are initialized to the offline weights every time this mode starts, get updated for each frame. During the online mode, if the RMSE value ($RMSE_n^{on}$) violates $RMSE_{Th}$, a validation step with a predetermined patience value ($P$) is performed based on the offline weights. If the comparison of $RMSE_n^{on}$ with $RMSE_n^{of}$ indicates that the online training process has reduced the accuracy in comparison to the offline trained model, a counter variable that holds the number of violated frames ($n_v$) is increased by one. If $n_v$ reaches the predetermined value of $P$, the model switches to offline mode which means the weights are reset back to the values obtained by the offline training and the online training stops temporarily, until $RMSE_n^{of}$ violates $RMSE_{Th}$ again and the need for using the online training arises. The mode decision algorithm which handles the switching between the online and offline modes and invokes the online training phase, is described in Algorithm 1, with $N$ being the total number of frames in the current rendering session.

### 3.3. Benchmark notes

GAMORRA's benchmark suite also operates at an API level. This is the only component in GAMORRA that interacts with the underlying hardware and the graphics driver. Since the graphics API's structure is not affected by the updates that the graphics driver receives, GAMORRA needs to rerun the benchmarks to capture the performance changes that these new functionalities cause. Hence, the core model of GAMORRA needs no changes in such scenarios and is the same for all the GPUs that use a certain graphics API. To perform the benchmarks and obtain the performance function of each stage of the pipeline, the number of input elements and the number of assembly operations for shaders is increased gradually until an upper bound is reached. The theoretical upper bound on the number of input elements is defined as the *resource limits* of Direct3D 11 [32]. However, running the benchmarks with the number of inputs close to the resource limits causes the frametimes to be much higher than the normal and acceptable values in modern applications (e.g. 30 fps). As a trade-off between frame rate generality and benchmark speed, 100 ms (or 10 fps) is chosen as the maximum acceptable frametime (or minimum acceptable frame rate) in the benchmarks. By choosing 100 ms, the benchmark would not run for the input load values that cause the frame rate to drop under 10 fps which effectively improves the benchmarking time. Since 10 fps is a fairly low frame rate for modern real-time applications, 100 ms is a reasonable choice in terms of generality.

The benchmarks to obtain the time complexity of IL shader assembly operators are designed such that only one opcode (e.g., add) is tested in each benchmark for a certain number of iterations. The recorded time for each benchmark is then divided by the number of iterations to obtain a rough estimate of the time complexity for the tested operator and stage, i.e., $OP_j$ in (6).

In addition to each stage's functionality, other pipeline states, such as the presentation model or blend mode, directly affect the final performance and need to be addressed in the benchmarks. For example, in Direct3D 11, DXGI_SWAP_EFFECT_FLIP_SEQUENTIAL and DXGI_SWAP_EFFECT_DISCARD presentation models would give two very different performance results based on the rendering resolution, which need to be taken into account while designing the benchmarks.

Table 1: Model parameters and important notes

| Stage | Model parameter | Notes |
|---|---|---|
| **IA** | #Vtx | VS, HS, DS, GS, PS set to pass through, No rasterization |
| **VS** | #Vtx, #Ops | HS, DS, GS, PS set to pass through, No rasterization |
| **HS** | #Vtx, #Ops | VS, DS, GS, PS set to pass through, No rasterization |
| **Tessellation** | #Patches, #Tess | VS, DS, GS, PS set to pass through, No rasterization |
| **DS** | #Vtx, #Ops | VS, GS, PS set to pass through, No rasterization |
| **GS** | #Vtx, #Ops | VS, HS, DS, PS set to pass through, No rasterization |
| **Rasterizer** | #Frg | VS, HS, DS, GS, PS set to pass through, 4 vertices required to map a texture for rasterization |
| **PS** | #Frg, #Ops | VS, HS, DS, GS set to pass through, 4 vertices required to map a texture for rasterization |
| **OM** | #Frg | VS, HS, DS, GS, PS set to pass through, 4 vertices required to map a texture in each layer for rasterization |
| **CS** | #Element | VS, HS, DS, GS, PS set to pass through, No rasterization |

Table 2: Sequence characteristics for the experimented games

| Game | Abbrv. | Genre | Resolution | S1 Average frametime (ms) | S1 Standard deviation (ms) | S2 Average frametime (ms) | S2 Standard deviation (ms) |
|---|---|---|---|---|---|---|---|
| **Bad Company 2 (2010)** | BC2 | FPS | 1280x720 | 12.22 | 0.943 | 07.47 | 0.624 |
| **Dirt 3 (2011)** | D3 | Racing | 1280x720 | 07.97 | 0.588 | 04.78 | 0.298 |
| **Far Cry 3 (2012)** | FC3 | FPS | 1920x1080 | 24.51 | 1.213 | 10.27 | 0.743 |
| **Rocket League (2015)** | RL | Racing/sport | 1280x720 | 16.13 | 1.247 | 09.76 | 0.987 |
| **Splinter Cell (2013)** | SC | Third Person | 1280x720 | 14.74 | 3.093 | 07.90 | 2.253 |
| **Trine 4 (2019)** | T4 | Side scroller | 1920x1080 | 16.42 | 2.582 | 09.45 | 2.012 |
| **Sniper Elite 4 (2017)** | SE4 | FPS | 1280x720 | 16.78 | 2.056 | 10.66 | 1.596 |
| **Mortal Shell (2020)** | MSh | Third Person | 1280x720 | 17.95 | 2.485 | 07.33 | 1.825 |
| **Fast and Furious (2021)** | FF | Racing | 1920x1080 | 16.37 | 1.991 | 08.28 | 1.371 |

Table 3: System configuration

| | S1 | S2 |
|---|---|---|
| **CPU** | i7-7500U | i5-7300HQ |
| **GPU** | 950m 2GB | RX 560 2GB |
| **RAM** | 8 GB DDR4 1200MHz | 8 GB DDR4 1200MHz |

Table 4: Training configuration

| | Offline | Online |
|---|---|---|
| **Initial LR** | 0.01 | 0.01 |
| **#Samples** | 720*#DC | #DC |
| **Batch size** | 32 | #DC |
| **#Epochs** | 200 | 1 |
| **Train/Test** | 0.3 | - |
| **Patience** | 10 | 10 |
| **RMSE$_{Th}$** | 0.5 | 0.5 |

Also, the benchmarks should account for techniques such as the post transform cache [33] and the early-z implemented in GPUs. The post transform cache causes the VS to be invoked less frequently for indexed draw calls. Hence, to properly have a 1:1 relation between the vertices and the number of times that a VS is invoked, indexed draws should be avoided unless the goal is to benchmark the post transform cache performance. The early-z test, also referred to as the early fragment test, depends on the functionality of the graphics driver and the GPU itself and they are not controlled explicitly by the API commands in Direct3D. Hence, a depth-only benchmark is used to determine the performance of the early-z process.

Table 1 shows a summary of all the requirements for each stage of the pipeline. The model parameters are the number of vertices (#Vtx), the number of assembly operations (#Ops), the number of patches (#Patch), the number of tessellations (#Tess), the number of primitives (#Prim), the number of target pixels or resolution (Res), the number of fragments (#Frg) and the number of input elements (#Element). It should be noted that for the Rasterizer and its following stages, the rasterization should be performed. Hence, to map a rectangular texture to the screen, at least 4 vertices are required. Also, the Res value determines the maximum resource resolution (texture, depth buffer, stencil buffer and etc.) of the current batch.

## 4. IMPLEMENATION AND RESULTS

APITrace [34] is used to intercept API commands that are produced by the rendering engine. Since computer games are the most computationally intensive applications that use rasterization-based graphics APIs to their maximum capacity, all the tests are done on modern AAA games. Also, the proposed model and all the mathematical calculations, including the matrix multiplications, are implemented using Tensorflow-GPU [35]. Nine computer games were chosen for the tests, namely, Dirt 3 (D3), Splinter Cell: Blacklist (SC), Battlefield Bad Company 2 (BC2), Far Cry 3 (FC3), Rocket League (RL), Trine 4 (RL), Sniper Elite 4 (SE4), Mortal Shell (MSh), and Fast and Furious Spy Racers: Rise of SH1FT3R (FF). Table 2 shows the game sequence characteristics such as genre, resolution, average frametime and the standard deviation of frametimes on two tested rendering systems. Since GAMORRA is designed to be independent of the underlying hardware, it is mandatory to perform the experiments on more than one rendering platform. The configurations of these tested devices are listed in Table 3.

The configurations of both training phases, which were determined through experimentation, are reported in Table 4. The offline training process uses the data of 720 frames of different sections of each level to train the model based on the graphics data of that level. Hence, the total number of samples would equal the average number of drawcalls (#DC) in a frame, times the number of frames. Also, the train-test ratio (Train/Test)

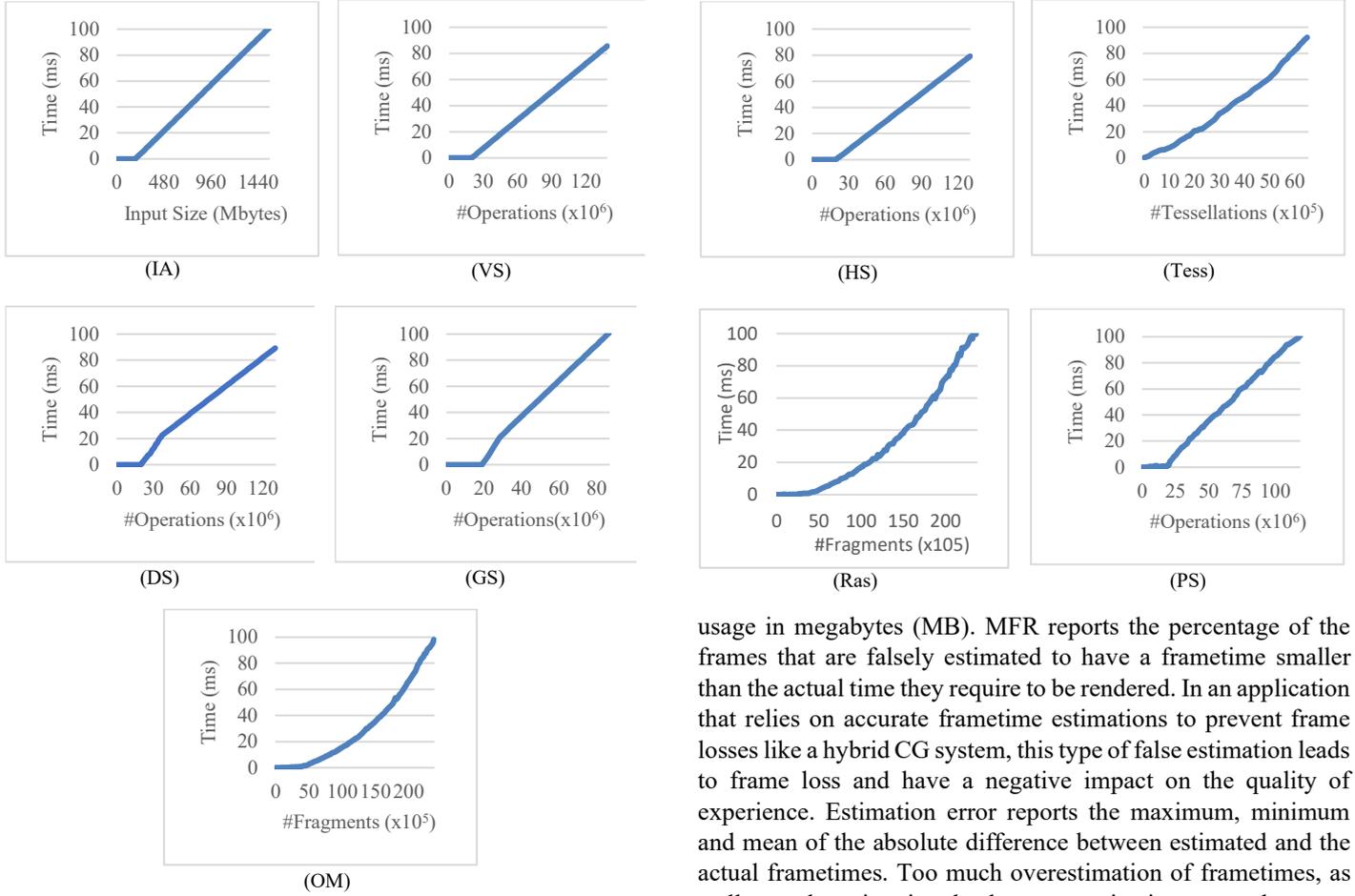

Figure 3 The performance charts of IA, VS, HS, Tessellation, DS, GS, Rasterization, PS and OM stages

of the offline training is set to 0.3.

The proposed method in this paper focuses on GPUs. However, the main CPU also affects frametimes immensely and can potentially become the bottleneck in a rendering session. Using APITrace makes sure that a lot of CPU intensive tasks like processing the IA is already carried out and the CPU is only dedicated to render-related processes. Many CPU performance models have been discussed in the literature [36] from as late as 1977 and is much simpler to obtain. For the purposes of this paper, the proposed method in [37] is used for CPU time complexity modeling. This model uses a deep neural network specifically for Intel CPUs based on available benchmarks such as the SPEC CPU2006 and Geekbench 3.

Three methods are chosen to compare to GAMORRA. These methods are the AutoRegressive (AR) moving average [9], Frame Complexity Model (FCM) [23], and a frequency-based method (FRQ) [10]. All these methods are independent of the underlying architecture and do not require any changes to the target software or hardware. Five metrics are considered in the experiments: Missed frames ratio (MFR) in percentage (%), frametime estimation error in milliseconds (ms), time complexity (ms), frametime overhead (%), and system memory usage in megabytes (MB). MFR reports the percentage of the frames that are falsely estimated to have a frametime smaller than the actual time they require to be rendered. In an application that relies on accurate frametime estimations to prevent frame losses like a hybrid CG system, this type of false estimation leads to frame loss and have a negative impact on the quality of experience. Estimation error reports the maximum, minimum and mean of the absolute difference between estimated and the actual frametimes. Too much overestimation of frametimes, as well as underestimation, leads to a negative impact on the target application's performance and should be accounted for in the experiments. Time complexity determines the amount of time overhead that GAMORRA forces upon the system which is mostly comprised of the MLR's parameter estimation processing time. The frametime overhead determines the percentage of the overall frametime that is due to GAMORRA's time complexity. And finally, system memory usage determines the amount of RAM in megabytes, required to run the model.

It should be noted that AR requires to consider the frametimes of a certain number of previous frames to predict the current frametime, referred to as the sequence length, which directly affects the performance of the model. The sequence length of AR is set to 10 in the experiments, which is reported to yield the best results in [9]. Also, FCM uses three weight coefficients for the number of vertices, the number of textures and the number of commands which are all set to $1/3$ as per FCM's original design [23]. And finally, FRQ [10] does not require any specific configurations for the experiments.

Before experimenting on the games, the performance function of the target hardware should be known.

### 4.1. Obtaining performance functions and model parameters

The results of benchmarking S1 are shown in Figure 3. With a $\beta_0$ of 6.966 ms, Figure 3 (IA) shows the result of the performance benchmark of the IA stage. Since loading about 1500 MBs of vertex data causes the pipeline to take more than 100 ms, it is evident that GTX950m is struggling to load them.

Table 5: Active and inactive shaders for the tested games

| Game | VS | Tess. | GS | PS | CS |
|---|---|---|---|---|---|
| BC2 | ✓ | ✗ | ✗ | ✓ | ✗ |
| D3 | ✓ | ✓ | ✗ | ✓ | ✓ |
| FC3 | ✓ | ✓ | ✓ | ✓ | ✓ |
| RL | ✓ | ✓ | ✗ | ✓ | ✓ |
| SC | ✓ | ✓ | ✗ | ✓ | ✗ |
| T4 | ✓ | ✗ | ✗ | ✓ | ✓ |
| SE4 | ✓ | ✓ | ✗ | ✓ | ✓ |
| MSh | ✓ | ✓ | ✓ | ✓ | ✓ |
| FF | ✓ | ✗ | ✓ | ✓ | ✓ |

Table 6: The values of some of the model parameters obtained from tested game sequences

| Game | #Vtx | | | #Attr | | | #Ops | | |
|---|---|---|---|---|---|---|---|---|---|
| | Min | Max | Mean | Min | Max | Mean | Min | Max | Mean |
| BC2 | 8 | 2134 | 752 | 1 | 7 | 3.5 | 7 | 192 | 32.7 |
| D3 | 4 | 2066 | 615 | 1 | 7 | 1.8 | 5 | 204 | 45.1 |
| FC3 | 4 | 2796 | 802 | 1 | 8 | 3.9 | 9 | 217 | 49.3 |
| RL | 4 | 1954 | 945 | 1 | 9 | 6.4 | 8 | 269 | 57.8 |
| SC | 8 | 2378 | 743 | 1 | 9 | 7.3 | 5 | 187 | 37.2 |
| T4 | 16 | 3072 | 1040 | 1 | 9 | 5.0 | 6 | 228 | 41.5 |
| SE4 | 12 | 3198 | 1816 | 1 | 13 | 3.7 | 7 | 236 | 31.4 |
| MSh | 12 | 3564 | 1792 | 1 | 16 | 10.4 | 9 | 253 | 48.3 |
| FF | 12 | 2974 | 1605 | 1 | 16 | 9.8 | 7 | 291 | 52.7 |

Table 7: The experimental results on MFR (%)

| Game | S1 | | | | S2 | | | |
|---|---|---|---|---|---|---|---|---|
| | AR | FCM | FRQ | GM | AR | FCM | FRQ | GM |
| BC2 | 11.54 | 3.08 | 12.32 | **2.35** | 11.39 | 3.21 | 11.42 | **2.46** |
| D3 | 08.79 | 3.69 | 09.35 | **2.16** | 09.76 | 3.62 | 09.16 | **2.33** |
| FC3 | 10.38 | 4.53 | 11.47 | **3.07** | 10.16 | 2.94 | 10.76 | **2.91** |
| RL | 09.21 | 3.24 | 08.44 | **2.58** | 10.09 | 2.86 | 09.49 | **2.27** |
| SC | 12.06 | 5.22 | 12.25 | **3.90** | 11.69 | 4.90 | 11.83 | **3.26** |
| T4 | 10.94 | 3.27 | 11.16 | **2.83** | 09.83 | 2.68 | 11.78 | **2.69** |
| SE4 | 11.92 | 4.75 | 12.13 | **3.24** | 12.78 | 5.11 | 12.66 | **2.70** |
| MSh | 10.87 | 4.20 | 11.04 | **3.06** | 11.80 | 4.28 | 11.47 | **2.72** |
| FF | 09.32 | 4.68 | 10.51 | **3.11** | 09.17 | 4.06 | 10.42 | **2.84** |

After being prepared by the IA stage, the vertices go through the VS stage. Tens of opcodes are available to be used by developers at this stage and all the other programmable stages and all of them should be benchmarked. As an example, the performance chart of the add operator of the VS stage is shown in Figure 3 (VS). This benchmark shows that a VS can perform of up to 130 million Add operations under a 100-ms time interval. Normally, the performance function of this stage would be a 3D chart, but, for simplicity and more comprehensible output, the number of attributes and the number of vertices is fixed and set to 1. For illustration purposes, the only attribute considered in the test in Figure 3 (VS), is position which is a 3-component floating point variable.

The results for the Tessellation stages are depicted in Figure 3 (HS), (Tess) and (DS). The results of the HS and DS stages for add operator are more similar to the VS stage in comparison to the Tessellator stage which is a fixed-function stage. For the Tessellator stage, level 3 tessellation was used and a total of $6.5*10^6$ tessellations make the rendering time to surpass 100 ms as depicted in Figure 3 (Tess).

The performance chart of the GS stage for add operator is depicted in Figure 3 (GS). This stage is also somewhat similar to the VS stage, albeit more demanding due to the fact that it operates on whole primitives instead of individual vertices.

Figure 3 (Ras) shows the benchmark results for the Rasterizer. As the number of fragments increase over 20 million, the Rasterizer starts to impose larger overhead on the pipeline and the GTX950m starts to struggle with the rasterization process. When the rasterization benchmark is 3 million pixels short of the 30 million, the rendering time gets closer enough to the unacceptable 100 ms.

The result of the PS stage benchmark for add operator is depicted in Figure 3 (PS). Performing $10^8$ operations would practically increase the processing time of the pipeline to well over 100 ms.

Finally, the performance chart of the OM stage is depicted in Figure 3 (OM). The load size of the OM stage is highly dependent upon the number of fragments and reaching $25*10^6$ would violate the 100-ms limit that is chosen for the benchmarks.

These benchmarks show a roughly linear relation between the load of the pipeline and the processing time that this load imposes on the system. If the input load is light enough (e.g., less than 5 million pixels and 10 million primitives in the case of rasterization), then the overall processing time of the pipeline would be roughly close to the value of $\beta_0$. So, it is safe to assume that an MLR-based model is flexible enough to model an API's pipeline and as will be discussed later on, it is not too complicated and computationally expensive to have a negative impact on the performance of the system.

If any of the tested games utilizes any of the optional stages, they should be also taken into account. Table 5 represents the active (✓) and pass-through or inactive (✗) shaders for the tested games with VS and PS always active for all the games. Also, some of the important graphics characteristics of the tested games are represented in Table 6. The number of vertices (#Vtx) determines the minimum, maximum and average values of the number of input vertices to a VS. The number of attributes (#Attr.) also determines the minimum, maximum and average values of the number of attributes of vertices. The minimum, maximum and average number of operations (#Ops) in shaders are also reported in Table 6.

### 4.2. Estimating frametimes

After establishing the performance functions of the target graphics card and training the model parameters, GAMORRA is ready to be used to estimate frametimes. To experiment on the games, a 5-minute gameplay sequence of each game was recorded using API Trace so it can be replayed multiple times to replicate the exact sequence of the gameplay.

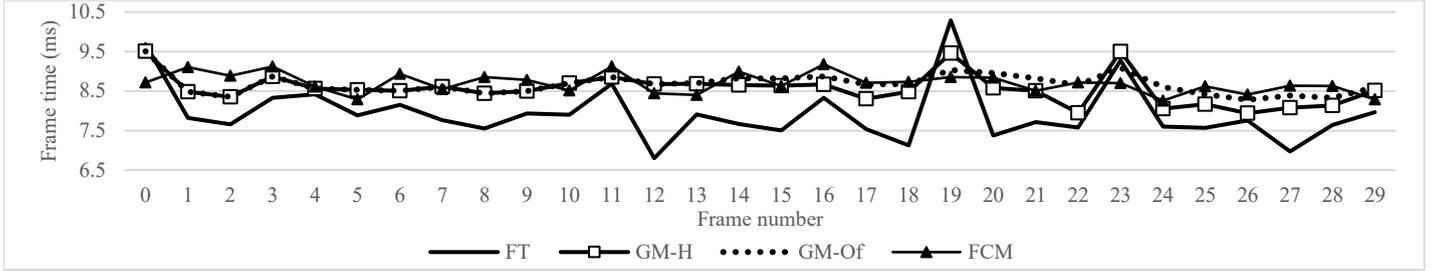

Figure 4 Frametime estimation results for D3 on S1

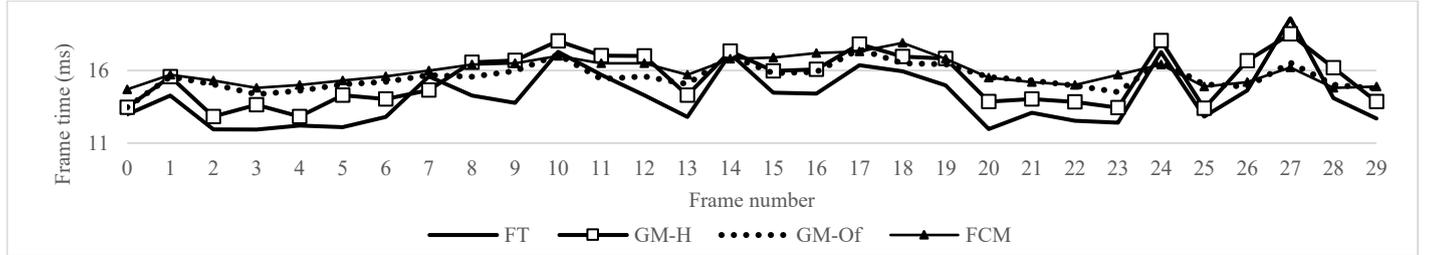

Figure 5 Frametime estimation results for SC on S1

### 4.2.1. Missed Frames Ratio (MFR)

Table 7 shows the MFR values for each method and each game. Higher MFR values indicate more frametime underestimations which can negatively impact the quality of experience of frame loss sensitive applications such as computer games [38]. Table 7 compares the result of GAMORRA (GM) with AR, FCM, and FRQ. The results show that GAMORRA outperforms all three methods from an MFR perspective on both S1 and S2 platforms. Among the tested methods, GAMORRA has the lowest MFR values with an average of 2.80% followed by FCM, AR, and FRQ with average miss rate values of 3.91%, 10.65%, and 10.98%, for both S1 and S2 platforms. FCM which takes a more fine-grained approach compared to AR and FRQ that rely only on the previous frames' rendering times, demonstrates closer results to GAMORRA. This is due to the fact that FCM considers the frame structure in the form the number of drawcalls, the texture size and the number of vertices which is still a coarse-grained approach compared to GAMORRA. Hence, it is outperformed by the proposed model. Moreover, FCM neither uses an online training method nor a comprehensive benchmark like GAMORRA does. This puts FCM at more of a disadvantage which leads to the loss of accuracy that is evident in Table 7.

Figure 4 and Figure 5 show the results of frametime estimation for D3 and SC on S1 in comparison to the actual frametimes (FT) in a 30-frame sequence of gameplay for each game. GAMORRA is tested for both offline-only (GM-Of) and hybrid (GM-H) training configurations to demonstrate the impact of the online training scheme. D3 and SC are chosen as the best-case and worst-case scenarios since they have the lowest and highest MFR values, respectively. To keep the charts simple, AR and FRQ methods are discarded in these figures.

D3 utilizes Ego engine [39] which is specifically designed and used for racing games, thus, performs efficiently on a mid-range GPU. On the other hand, SC uses a pretty much outdated unreal engine 2.5 [40] which is pushed to its limits for the best visual output leading to a somewhat unstable rendering performance. Hence, although GAMORRA outperforms FCM, still, a drop in accuracy is experienced for both GM-H and GM-Of. Figure 4 shows that GM-H excels at modeling the rendering times in comparison to FCM that does not utilize a hybrid training scheme. In this chart, GAMORRA switches to online training on frame 13 which leads to a divergence in GM-H and GM-Of estimations. It is evident on frames 19 and 23 that GM-Of fails to react in time to frametime variations that are caused by changes other than workload variations. This also applies to FCM as well.

Figure 5 shows the frametime results for SC. This chart shows that GM-H reacts faster to the fluctuations in frametimes which is a result of both the more detailed core model and the more sophisticated training scheme in comparison to FCM. In this chart, the system switches to online training on the second frame. FCM and GM-Of react to workload variations to some extent, like frame 10. However, GM-H can mimic more aggressive variations better than GM-Of and FCM, like frames 24 and 27. Using frame workload structure in conjunction with a hybrid training scheme results in faster and more accurate reactions to frametime variations that are caused by both workload variations and changes in runtime conditions.

### 4.2.2. Estimation error

The results of the estimation error are reported for each of the tested methods for both tested platforms in Table 8. Also, the percentage of the estimation error is reported in Table 9. The percentage of estimation error is highly dependent upon the value of frame rate and as the frame rate increases, this percentage increases proportionally.

The average estimation error reported in Table 8 shows the dominance of GM-H for all the tested games on both S1 and S2 platforms. On S1, GM-H has an average of 4.13% less estimation error for the tested games compared to FCM. This value for S2 drops to 3.47%. This indicates that GM-H is capable of handling a less stable rendering session on weaker devices such as S1 compared to a more powerful machine like S2. The estimation error gap between GM-H and the other two methods on S1 equals 9.74% and 14.79% for AR and FRQ, respectively. These values on S2 drop to 9.57% and 13.5% for AR and FRQ, respectively. Although GM-Of does not take advantage of GAMORRA's hybrid training scheme, it still manages to outperform all the other methods other than GM-H. GM-Of is outperformed by GM-H by 1.81% and 1.64% on S1 and S2, respectively. The larger estimation gap between GM-H and MG-Of on S2 compared to S1 indicates that GM-H's online training scheme is responsible for the larger estimation gap between GM-H and the other models that do not utilize an online training method.

Table 8: The experimental results on estimation error (ms)

| Game | S1 | | | | | S2 | | | | |
|------|------|------|------|-------|-------|------|------|------|-------|-------|
|      | AR   | FCM  | FRQ  | GM-H  | GM-Of | AR   | FCM  | FRQ  | GM-H  | GM-Of |
| BC2  | 2.425 | 1.941 | 3.847 | **1.248** | 1.495 | 2.375 | 1.713 | 2.906 | **1.093** | 1.217 |
| D3   | 0.810 | 0.682 | 1.614 | **0.576** | 0.638 | 0.752 | 0.626 | 1.292 | **0.45** | 0.583 |
| FC3  | 3.167 | 1.859 | 3.991 | **1.108** | 1.793 | 2.954 | 1.451 | 3.293 | **0.801** | 1.141 |
| RL   | 1.403 | 1.316 | 2.753 | **1.017** | 1.211 | 1.312 | 1.235 | 2.626 | **0.743** | 1.094 |
| SC   | 4.104 | 2.991 | 4.280 | **1.564** | 2.074 | 3.761 | 2.021 | 4.086 | **1.276** | 1.569 |
| T4   | 1.938 | 1.853 | 2.494 | **1.460** | 1.798 | 1.828 | 1.420 | 2.132 | **1.143** | 1.31 |
| SE4  | 5.742 | 3.076 | 6.059 | **1.953** | 2.231 | 5.390 | 2.708 | 5.597 | **1.621** | 1.917 |
| MSh  | 3.906 | 2.244 | 4.271 | **1.786** | 2.002 | 3.158 | 1.870 | 4.034 | **1.463** | 1.866 |
| FF   | 3.436 | 2.473 | 4.005 | **1.831** | 2.013 | 2.782 | 2.146 | 3.519 | **1.695** | 1.954 |

Table 9: The experimental results on estimation error (%)

| Game | S1 | | | | | S2 | | | | |
|------|------|------|------|-------|-------|------|------|------|-------|-------|
|      | AR   | FCM  | FRQ  | GM-H  | GM-Of | AR   | FCM  | FRQ  | GM-H  | GM-Of |
| BC2  | 19.84 | 15.88 | 31.48 | **10.21** | 12.23 | 31.79 | 22.93 | 38.9 | **14.63** | 16.29 |
| D3   | 10.16 | 8.56 | 20.25 | **7.23** | 8.01 | 15.73 | 13.1 | 27.03 | **9.41** | 12.2 |
| FC3  | 12.92 | 7.58 | 16.28 | **4.52** | 7.32 | 28.76 | 14.13 | 32.06 | **7.8** | 11.11 |
| RL   | 8.7 | 8.16 | 17.07 | **6.31** | 7.51 | 13.44 | 12.65 | 26.91 | **7.61** | 11.21 |
| SC   | 27.84 | 20.29 | 29.04 | **10.61** | 14.07 | 47.61 | 25.58 | 51.72 | **16.15** | 19.86 |
| T4   | 11.8 | 11.29 | 15.19 | **8.89** | 10.95 | 19.34 | 15.03 | 22.56 | **12.1** | 13.86 |
| SE4  | 34.22 | 18.33 | 36.11 | **11.64** | 13.3 | 50.56 | 25.4 | 52.5 | **15.21** | 17.98 |
| MSh  | 21.76 | 12.5 | 23.79 | **9.95** | 11.15 | 43.08 | 25.51 | 55.03 | **19.96** | 25.46 |
| FF   | 20.99 | 15.11 | 24.47 | **11.19** | 12.3 | 33.6 | 25.92 | 42.5 | **20.47** | 23.6 |

Table 10: The experimental results on average time complexity (ms)

| Game | S1 | | | | S2 | | | |
|------|------|------|------|------|------|------|------|------|
|      | AR   | FCM  | FRQ  | GM   | AR   | FCM  | FRQ  | GM   |
| BC2  | 1.63 | 1.54 | **1.47** | 2.15 | 0.94 | 0.66 | **0.59** | 1.23 |
| D3   | 1.37 | 1.17 | **1.01** | 2.02 | 0.72 | 0.47 | **0.40** | 1.18 |
| FC3  | 1.46 | 1.19 | **1.08** | 1.79 | 0.64 | 0.52 | **0.43** | 1.03 |
| RL   | 1.54 | 1.26 | **1.10** | 2.11 | 0.86 | 0.68 | **0.53** | 1.31 |
| SC   | 1.82 | 1.74 | **1.65** | 1.83 | 1.02 | 0.83 | **0.67** | 1.16 |
| T4   | 1.70 | 1.41 | **1.29** | 2.04 | 0.81 | 0.59 | **0.46** | 1.22 |
| SE4  | 1.51 | 1.43 | **1.24** | 1.99 | 0.62 | 0.52 | **0.45** | 1.19 |
| MSh  | 1.67 | 1.41 | **1.35** | 2.03 | 1.02 | 0.61 | **0.49** | 1.20 |
| FF   | 1.49 | 1.18 | **1.01** | 1.67 | 1.01 | 0.44 | **0.40** | 1.01 |

Table 11: The experimental results on estimation overhead (%)

| Game | S1 | | | | S2 | | | |
|------|------|------|------|------|------|------|------|------|
|      | AR   | FCM  | FRQ  | GM   | AR   | FCM  | FRQ  | GM   |
| BC2  | 11.769 | 11.192 | **10.738** | 14.962 | 11.177 | 8.118 | **07.32** | 14.138 |
| D3   | 14.668 | 12.801 | **11.247** | 20.220 | 13.091 | 8.952 | **7.722** | 19.799 |
| FC3  | 05.622 | 04.630 | **04.220** | 06.806 | 05.866 | 4.819 | **4.019** | 09.115 |
| RL   | 08.715 | 07.246 | **06.384** | 11.568 | 08.098 | 6.513 | **5.151** | 11.834 |
| SC   | 10.990 | 10.558 | **10.067** | 11.044 | 11.435 | 9.507 | **7.818** | 12.804 |
| T4   | 09.382 | 07.908 | **07.284** | 11.051 | 07.895 | 5.876 | **4.642** | 11.434 |
| SE4  | 08.256 | 07.853 | **06.881** | 10.602 | 05.496 | 4.651 | **4.050** | 10.042 |
| MSh  | 08.512 | 07.283 | **06.995** | 10.160 | 12.216 | 7.683 | **6.266** | 14.068 |
| FF   | 08.343 | 06.724 | **05.811** | 09.257 | 10.872 | 5.046 | **4.608** | 10.872 |

### 4.2.3. Time complexity and overhead

The time complexity of the tested methods is reported in Table 10. FRQ only relies on the previous frametimes and the GPU frequency in a simple linear method which leads to the lowest complexity among the tested methods. FCM, which uses a scheme based on frame structure similar to GAMORRA, yet with a simpler model, has lower time complexity than the proposed model. This is due to the simpler nature of its core model and the absence of any online training scheme. AR relies on an iterative process to obtain its model parameters. Hence, it is expected to have a larger time complexity compared to the other two methods. GAMORRA has the largest time complexity among the tested models. In addition to its more complex core model, the online training scheme of the proposed model is expected to have a slight negative impact on its time complexity. However, although GAMORRA has a larger time complexity compared to the other tested methods, it still manages to perform acceptably and in real-time, especially considering its superior accuracy.

Table 11 shows the overhead of GAMORRA in comparison to the three tested methods for both S1 and S2 platforms. As expected from the time complexity results represented in Table 10, FRQ which is the least computationally complex method among all the tested methods, outperforms the other methods from an estimation overhead standpoint. An average overhead of 7.74%, 8.47%, 9.58% and 11.74% is experienced for FRQ, FCM, AR and GM, respectively, on S1, and 5.73%, 6.80%, 9.57% and 12.68% for FRQ, FCM, AR and GM, respectively, on S2. The results of S1 are mostly similar to the results of S2. However, there are cases in which there is a wider gap between the overhead of S1 and S2, which indicates that extra processing power of S2 has a larger impact on the performance of the target game than it has on the performance of the workload model. Since a large portion of GAMORRA's time complexity is caused by a serial analysis of the shader code complexity, the overhead does not scale with the processing power of the target device as much as the other methods. This also applies to AR which uses an iterative parameter estimation method.

It should be noted that these games are designed to be rendered in 60 fps and the overhead is highly affected by the value of the frametimes. As the average frametime increases, the overhead would decrease. For example, if the experimented game runs at 30 fps and the frame rate is not capped artificially, each frame would take about 33 ms to be rendered. Hence, the overhead values would almost be cut in half.

### 4.3. System resource usage

GAMORRA's implementation is comprised of two distinct parts: the analyzer and the core model. The analyzer analyzes the intercepted graphics API commands and stores the required data (e.g., pointers and properties of buffers, textures, and resource views like size, resolution and usage). After analyzing the graphics data, the core model is fed with the required input parameters to be trained and used to predict frametimes.

Table 12 reports the system memory usage of all the tested methods along with the analyzers of GAMORRA (GMA) and FCM (FCMA) in megabytes (MB). FRQ which is the simplest model among the tested models, requires at most 1.0 MB of system memory. FCM also requires a frame analyzer similar to GAMORRA, albeit much simpler, as is evident from the results in Table 12. FCMA requires 15.67 MB of memory on average while GMA which needs to consider more parameters than FCMA requires an average of 56.78 MB of system memory. While neural networks are known to demand large amount of system memory as their number of layers increase, the choice of MLR for the core model of GAMORRA ensures a limited and small amount of memory requirement while it uses 32-bit floating point precision for its weights. The reported results in Table 12 shows that the memory usage of the core model of GAMORRA will not surpass 3 MB at most. FCM uses a simpler core model compared to GAMORRA which has led to 0.87 MB less memory usage on average. Since AR only needs to consider the frametime of a limited number of previous frames, it does not need to use an analyzer like GAMORRA and FCM. The memory usage of the AR is directly impacted by the sequence length of the model. With the sequence length set to 10, the AR's memory usage is almost similar to GAMORRA. It should be noted that the training session's memory usage is different from the memory usage of the model during prediction. For both AR and GAMORRA (offline), the training session's memory is almost similar and it is equal to 1.32 gigabytes on average.

Table 12: System memory usage (MB) of the analyzer and the core model of GAMORRA in comparison to AR, FCM, and FRQ

| Game | AR | FCM | FCMA | FRQ | GM | GMA |
|---|---|---|---|---|---|---|
| BC2 | 0.9 | 1.4 | 18 | 0.7 | 2.4 | 64 |
| D3 | 1.2 | 1.5 | 13 | 0.9 | 2.1 | 46 |
| FC3 | 1.4 | 1.8 | 19 | 1.0 | 2.5 | 67 |
| RL | 0.8 | 1.3 | 10 | 0.5 | 2.3 | 29 |
| SC | 1.1 | 1.7 | 15 | 0.8 | 2.5 | 48 |
| T4 | 0.6 | 1.2 | 16 | 0.4 | 2.2 | 62 |
| SE4 | 0.9 | 1.5 | 21 | 0.5 | 2.4 | 78 |
| MSh | 1.2 | 1.7 | 17 | 0.9 | 2.6 | 63 |
| FF | 1.0 | 1.6 | 12 | 0.7 | 2.5 | 54 |

### 4.4. Overall verdict

The experimental results from the previous subsections reveals that GAMORRA can outperform the other three tested methods in terms of accuracy. This comes at the cost of more time complexity and more memory usage compared to FRQ and FCM. However, the time complexity and the memory usage are reasonable for a real-time application and do not compromise the model's performance. For less varied graphical workloads, GAMORRA and FCM perform similarly with a slight edge in accuracy for the proposed model. As the graphical workload of the games and the system conditions become more varied, GAMORRA tends to provide better estimations compared to the other methods and adapts to frametime variations more accurately. The results for GAMORRA with an the offline-only training scheme indicates that it has some of the flaws of FCM in adapting to changes in system conditions during runtime. However, it still achieves less estimation error compared to FCM due to its detailed core model.

### 5. CONCLUDING REMARKS

This paper proposes GAMORRA, an API-level workload model for rasterization-based graphics pipeline architectures. Modeling the workload of a game's frames proves useful in different applications like DVFS-based power management schemes in smartphones or estimation of performance measures like frametimes in a graphics streaming-based CG system. The API-level approach lets GAMORRA to work without the need to modify the source code of the target application and the rendering hardware which is an essential part of previous studies to compensate for the lack of sufficient depth in their core model. To account for the high-level approach of GAMORRA and its lack of accuracy, an MLR-based core model, a hybrid online/offline training method to train the model and a benchmark suite to tune the model parameters according to the performance of the target rendering system are proposed. GAMORRA takes into account the overall structure of a graphics rendering pipeline using an MLR model along with the size of the input data, i.e., vertex numbers and texture resolution as the explanatory variables. Also, the complexity of each shader is taken into account as well. The experiments were performed on two different rendering platforms with three other workload models. The experimental results show a meaningful estimation error reduction in comparison to previously proposed methods while keeping the time complexity and the time overhead imposed on the system within an acceptable range. Also, the miss rate for underestimated frames is reduced significantly.